\documentclass{icrc2009}
\usepackage{amssymb}
\usepackage{graphicx}   
\usepackage[caption=false]{caption}    
\usepackage[font=footnotesize]{subfig} 
\usepackage{fixltx2e}
\usepackage{url}

\newcommand{\shorttitle}[1]%
{\markboth{Proceedings of the 31\MakeLowercase{$^{st}$} ICRC, {\L}\'{o}d\'{z} 2009}{#1} }
\newcommand{\etal}{\MakeLowercase{\textit{et al.}}} 

\begin{document}
\title{On the statistical effects of multiple reusing of simulated air showers in detector simulations}

\author{\IEEEauthorblockN{A.D. Supanitsky \IEEEauthorrefmark{1} and
			  G. Medina-Tanco \IEEEauthorrefmark{1}
}
\\
\IEEEauthorblockA{\IEEEauthorrefmark{1} Departamento de F\'isica de Altas Energ\'ias, Instituto de Ciencias 
Nucleares, Universidad Nacional Aut\'onoma \\ de M\'exico, A. P. 70-543, 04510, M\'exico, D. F., M\'exico.}}

\shorttitle{A.D. Supanitsky \etal Multiple reusing of simulated air showers}
\maketitle

\begin{abstract}

The simulations of extensive air showers as well as the detectors involved in their detection play a 
fundamental role in the study of the high energy cosmic rays. At the highest energies the detailed 
simulation of air showers is very costly in processing time and disk space due to the large number 
of secondary particles generated in interactions with the atmosphere, e.g. $\sim 10^{11}$ for 
$10^{20}$ eV proton shower. Therefore, in order to increase the statistics, it is quite common to 
recycle single showers many times to simulate the detector response. In this work we present a detailed 
study of the artificial effects introduced by the multiple use of single air showers for the detector 
simulations. In particular, we study the effects introduced by the repetitions in the kernel density 
estimators which are frequently used in composition studies.

\end{abstract}


\section{Introduction}

Air shower and detector simulations play a fundamental role in the study of cosmic rays. In particular,
arrays of surface detectors that do not have fluorescence telescopes to calibrate the energy scale, 
must resort to simulated data in order to estimate the energy of the primary particle. Furthermore, the 
primary mass is also obtained comparing experimental data with simulations.

There are several Monte Carlo programs for air shower simulation, the most used in the literature are 
AIRES \cite{AIRES}, CORSIKA \cite{CORSIKA}, and CONEX \cite{conex}, the latter for a fast simulation of the 
longitudinal shower development. Since the number of particles produced in a shower can be extremely large, e.g., 
$\sim 10^{11}$ for a $10^{20}$ eV proton shower, the computer processing time and disk space needed are also very 
large, even if unthinning methods \cite{Hillas:85,Hillas:97} are used. Due to this difficulty it is a common 
practice to reuse the same shower for generating several events (see for example 
\cite{Ave:02,SupaCompo:09}). This practice is more common in simulations that include surface 
detectors because, for fluorescence telescopes, very fast Monte Carlo programs like CONEX, have very fast and 
efficient algorithms for the generation of longitudinal profiles. 

In this work we present a study of the effects of using multiple repetitions of individual showers \cite{SupaRep:08}, 
applied to the simulation of detectors, on the evaluation of standard estimators of the expected value, variance, and 
covariance. We study in detail the effects introduced in the kernel density estimators, which are analytical estimates 
of the underlying distribution function obtained from a finite sample of events. In cosmic rays physics this technique 
is used mainly in connection with composition analyses (see for example \cite{Antoni:03,SupaCompo:09,SupaClass:09}); 
however, it is also extensively used in many different areas of knowledge \cite{Silvermann:86} to which this work can 
be directly extended.

\section{Analytical Treatment}

As mentioned in the introduction, we want to study the potential distortions introduced by reusing individual showers 
to maximize the statistics when simulating the response of a detector. Let us start with the optimum case in which each 
individual shower is used only once and, therefore, best reproduces reality. 

Let $\mathbf{y}$ be a $d$-dimensional vector composed by physical observables (e.g. mass sensitive parameters) 
distributed as $g(\mathbf{y})$ and let $\mathbf{z}$ be a random vector, distributed as $h(\mathbf{z})$, that takes 
into account the effects of the detectors and the corresponding reconstruction method such that, after measuring 
and reconstructing the empirical information, a vector $\mathbf{x}=\mathbf{y}+\mathbf{z}$ is obtained. The distribution 
function of $\mathbf{x}$ is the convolution of $g(\mathbf{y})$ and $h(\mathbf{z})$,
\begin{equation}
f(\mathbf{x}) = g\circ h(\mathbf{x}) = \int d\mathbf{y}\ g(\mathbf{y}) h(\mathbf{x}-\mathbf{y}). 
\label{fdef}
\end{equation}

Suppose that we have a sample of $N$ independent events of the distribution $f$, 
$\mathbf{x}_1 = \mathbf{y}_1 + \mathbf{z}_1, \ldots , \mathbf{x}_N = \mathbf{y}_N + \mathbf{z}_N.$
The probability of this configuration can be written as,
\begin{eqnarray}
P(\mathbf{y}_1 \ldots \mathbf{y}_N, \mathbf{z}_1 \ldots \mathbf{z}_N) &=& g(\mathbf{y}_1)\ldots g(\mathbf{y}_N)\times \nonumber \\ 
&& h(\mathbf{z}_1) \ldots h(\mathbf{z}_N), \nonumber \\ 
P(\mathbf{x}_1 \ldots \mathbf{x}_N) &=& f(\mathbf{x}_1) \ldots f(\mathbf{x}_N).
\label{Dsamp}
\end{eqnarray}

However, as previously noted, if single showers are recycled and used many times to simulate the response of the 
detectors, non-independent samples are obtained. If we use each shower of a sample of $M$ independent showers $m$ 
times to simulate the detectors  response, the following sample of size $N=M\times m$ is obtained, 
$\mathbf{x}_{11} = \mathbf{y}_1 + \mathbf{z}_{11}, \ldots,$ $\mathbf{x}_{1m} = \mathbf{y}_1 + \mathbf{z}_{1m}, \ldots,$
$\mathbf{x}_{M1} = \mathbf{y}_M + \mathbf{z}_{M1}, \ldots,$ $\mathbf{x}_{Mm} = \mathbf{y}_M + \mathbf{z}_{Mm}$,
where the notation used henceforth corresponds to $\xi^i_{\alpha a}$, where $i$ is the $ith$ coordinate of vector 
$\mathbf{\xi}$, $\alpha$ indicates the number of independent shower and $a$ the number of detector simulation 
performed using the $\alpha th$ shower. The probability of such configuration is given by
\begin{eqnarray}
P(\mathbf{y}_1 \ldots \mathbf{y}_N, \mathbf{z}_{11} \ldots \mathbf{z}_{Mm})%
&=& \prod_{\alpha=1}^M g(\mathbf{y}_\alpha) \prod_{a=1}^m h(\mathbf{z}_{\alpha a}) \nonumber \\
P(\mathbf{x}_{11} \ldots \mathbf{x}_{Mm}) &=& \prod_{\alpha=1}^M \int d\mathbf{y}_\alpha\ g(\mathbf{y}_\alpha)\times \nonumber \\
&&\prod_{a=1}^m h(\mathbf{x}_{\alpha a}-\mathbf{y}_{\alpha}).
\label{DsampRep}
\end{eqnarray}

\subsection{Mean, variance and covariance estimators}

Let us consider the average of the $ith$ coordinate of $\mathbf{x}$, $x^i$, for the realistic case
in which each shower is used only once to simulate the detector response,
\begin{equation}
\bar{x}^i = \frac{1}{N}\ \sum_{\alpha=1}^N x_\alpha^i.
\label{xav}
\end{equation}
By using Eq. (\ref{Dsamp}) it is easy to obtain the very well known expressions for the expected value and 
variance of $\bar{x}^i$,
\begin{eqnarray}
\label{Ex}
E[\bar{x}^i]&=&E[x^i] \\
\label{Vx}
Var[\bar{x}^i] &=& \frac{1}{N}\ Var[x^i].
\end{eqnarray}

The usual estimator of the covariance between two random variables is given by,
\begin{equation}
\label{CovEst}
\hat{C}_{ij} = \frac{1}{N-1}\ \sum_{\alpha=1}^N  (x_\alpha^i-\bar{x}^i) (x_\alpha^j-\bar{x}^j).
\end{equation}
For $i=j$ the estimator of the variance of $x^i$ is obtained, $s_i^2=\hat{C}_{ii}$. By using Eq. (\ref{Dsamp}) it can be 
shown that both estimators are non-biased,
\begin{eqnarray}
\label{Ecov}
E[\hat{C}_{ij}]&=&cov[x^i, x^j], \\
\label{Es}
E[s_{i}^2] &=& E[\hat{C}_{ii}]= Var[x^i].
\end{eqnarray}

For the case in which each shower is used several times to simulate the response of the detectors the average of $x^i$ 
is given by,
\begin{equation}
\bar{x}'^i = \frac{1}{M m}\ \sum_{\alpha=1}^M \sum_{a=1}^m x_{\alpha a}^i.
\label{AvRep}
\end{equation}

From Eqs. (\ref{DsampRep},\ref{AvRep}) it can be shown that,
\begin{eqnarray}
\label{Eavxrep}
E[\bar{x}'^i] &=& E[x^i], \\
\label{Vavxrep}
Var[\bar{x}'^i] &=& \frac{1}{Mm}\ Var[x^i]+\frac{m-1}{Mm}\ \int d\mathbf{y} d\mathbf{x}_1 d\mathbf{x}_2 \nonumber \\ 
&&(x^i_1-E[x^i])(x^i_2-E[x^i])\times \nonumber \\
&& g(\mathbf{y})\ h(\mathbf{x}_1-\mathbf{y})\ h(\mathbf{x}_2-\mathbf{y}),
\end{eqnarray}
which means that using samples obtained by reusing individual showers to simulate the detector 
response does not introduce any bias when calculating the average. However the fluctuations of 
$\bar{x}^i$ are increased by the generation of an additional term proportional to $(m-1)/Mm$. 

The estimator of the covariance, between $x^i$ and $x^j$, including multiple repetitions of the individual 
showers takes the form,
\begin{equation}
\hat{C}'_{ij} = \frac{1}{M m-1}\ \sum_{\alpha=1}^M \sum_{a=1}^m (x_{\alpha a}^i-\bar{x}'^i)%
(x_{\alpha a}^j-\bar{x}'^j).
\label{CovRep}
\end{equation}

The expected value of the covariance estimator is obtained from Eqs. (\ref{DsampRep}) and (\ref{CovRep}), 
\begin{eqnarray}
\label{Ecovxrep}
E[\hat{C}'_{ij}] &=& cov[x^i, x^j]-\frac{m-1}{Mm} \int d\mathbf{y} d\mathbf{x}_1 d\mathbf{x}_2 \nonumber \\ 
&& (x^i_1-E[x^i])(x^j_2-E[x^j]) g(\mathbf{y}) \times \nonumber \\
&& h(\mathbf{x}_1-\mathbf{y})\ h(\mathbf{x}_2-\mathbf{y}).
\end{eqnarray}
Therefore, as expected, the repetition of individual showers introduces a bias in the covariance estimator
because the events are not independent. The bias results proportional to $(m-1)/Mm$. 

As mentioned before, the expected value of the variance estimator is obtained setting $i=j$ in Eq. 
(\ref{Ecovxrep}),
\begin{eqnarray}
\label{Evarxrep}
E[{s'}_i^2] &=& Var[x^i]-\frac{m-1}{Mm} \int d\mathbf{y} d\mathbf{x}_1 d\mathbf{x}_2 (x^i_1- \nonumber \\
&&E[x^i]) (x^i_2-E[x^i])\ g(\mathbf{y}) \times \nonumber \\
&& h(\mathbf{x}_1-\mathbf{y})\ h(\mathbf{x}_2-\mathbf{y}),
\end{eqnarray}
which shows that also ${s'}_i^2$ is now a biased estimator of the variance of $x^i$.

\subsection{Density estimators}
\label{DensEst}

The density estimation technique consist in obtaining an estimator of the underlying density function from a given 
data sample \cite{Silvermann:86}. In one of the most widely used variants of that technique, 
a density estimator is obtained from a superposition of kernel functions centered at each event of the data sample. For 
$d$-dimensional data the kernel density estimator can be written as,
\begin{equation}
\label{Dest}
\hat{f}(\mathbf{x}) = \frac{1}{N}\ \sum_{\alpha = 1}^N \frac{1}{\sqrt{|H|}}\ K(H^{-1/2} \cdot (\mathbf{x}-\mathbf{x}_\alpha)),
\end{equation}
where $\mathbf{x}$ is a $d$-dimensional vector, $H$ is a symmetric, positively defined matrix (i.e., the symmetric, 
positively defined square-root matrix $H^{-1/2}$ exists) and $K(\mathbf{u})$ is the kernel function. The matrix $H$ 
gives the covariance between the different pairs of variables and also the degree of smoothing, i.e., the width of the 
kernel function.   

From Eqs. (\ref{Dsamp}) and (\ref{Dest}) the expected value of the density estimator is obtained,
\begin{equation}
\label{Bias}
E[\hat{f}(\mathbf{x})] = \frac{1}{\sqrt{|H|}}\ \int d\mathbf{x}' \ K(H^{-1/2} \cdot (\mathbf{x}-\mathbf{x}'))\ f(\mathbf{x}'),
\end{equation}
which shows that $\hat{f}(\mathbf{x})$ is a biased estimator of $f(\mathbf{x})$.

By using the Taylor expansion and retaining the dominant terms an approximated expression for the integrated 
mean square error $IMSE = \int d\mathbf{x}\ E[(\hat{f}(\mathbf{x})-f(\mathbf{x}))^2]$ is obtained,
\begin{eqnarray}
IMSE &\cong& \frac{h^4}{4}\ \int d\mathbf{x}\ \left[ \int d\mathbf{u}\ K(\mathbf{u}) \mathbf{u}^T V^{1/2}%
\times \right. \nonumber \\ 
&& D^2 f(\mathbf{x})\left. V^{1/2} \mathbf{u}  \right]^2 + \frac{R(K)}{N\ h^d\ \sqrt{|V|}},
\label{IMSEVarBiash}
\end{eqnarray}
where
\begin{equation}
[D^2f(\mathbf{x})]_{ij} = \frac{\partial^2 f}{\partial x^i \partial x^j}(\mathbf{x}) \ \ \ %
R(A) = \int d\mathbf{u}\ A^2(\mathbf{u}).
\label{D2f}
\end{equation}
Here we take $H^{-1/2}=V^{-1/2}/h$, where $h$ is a small parameter that parametrizes the degree of 
smoothing.

Minimizing $IMSE$ with respect to $h$, the well known expression of $h_{opt}$ is recovered,
\begin{equation}
\label{hopt}
h_{opt} \propto \frac{1}{N^{1/(d+4)}},
\end{equation}
where the constant of proportionality depends on $f(\mathbf{x})$, the unknown density function that we want to 
estimate.

Let us consider the case in which shower repetitions of individual showers are included. The density estimator
in this case is given by,
\begin{equation}
\label{DestRep}
\hat{f}'(\mathbf{x}) = \frac{1}{M m}\ \sum_{\alpha = 1}^M \sum_{a = 1}^m \frac{1}{\sqrt{|H|}}\ K(H^{-1/2} \cdot %
(\mathbf{x}-\mathbf{x}_{\alpha a})).
\end{equation}

It can be seen from Eqs. (\ref{DsampRep}) and (\ref{DestRep}), that the bias does not change when the repetitions are 
introduced. However, as expected, the variance increases,
\begin{eqnarray}
Var[\hat{f}'(\mathbf{x})]\ &\cong& \frac{1}{M m\ \sqrt{|H|}}\ R(K)\ f(\mathbf{x}) + \frac{m-1}{M m}  \nonumber \\ 
\label{VarRep}
&&\left( \int d\mathbf{y}\ g(\mathbf{y}) h^2(\mathbf{x}-\mathbf{y})- f^2(\mathbf{x})\right),\ \nonumber \\ 
\end{eqnarray}
where just the leading terms are retained. Consequently, the $IMSE$ takes in this particular case the form 
\begin{eqnarray}
IMSE' &\cong& \frac{h^4}{4}\ \int d\mathbf{x}\ \left[ \int d\mathbf{u}\ K(\mathbf{u}) \mathbf{u}^T V^{1/2} \times 
\right. \nonumber \\ 
&& \left. D^2 f(\mathbf{x}) V^{1/2} \mathbf{u}  \right]^2 + \frac{R(K)}{M m\ h^d\ \sqrt{|V|}} + \nonumber \\
&& \frac{m-1}{M m} \int d\mathbf{x} d\mathbf{y}\ g(\mathbf{y}) h^2(\mathbf{x}-\mathbf{y}) - \nonumber \\
&& \frac{m-1}{M m} R(f).
\label{IMSEVarBiashRep}
\end{eqnarray}

Eq. (\ref{IMSEVarBiashRep}) shows that the leading term introduced by the repetitions does not depend on $h$ and, 
therefore, the expression for $h_{opt}$ remains equal to the $m=1$ case. The only effect introduced by the repetitions 
of the individual showers is to increase the fluctuations of the estimator for each $\mathbf{x}$.

\section{Numerical Example}
\label{example}

In this section a numerical example that shows the predicted effects introduced by the shower repetitions is given. 
For that purpose, air showers simulations are performed using the program CONEX. A total of $N_{sh}=11000$ proton 
showers of primary energy $E=10^{19}$ eV and zenith angle $\theta = 30^\circ$ are generated.  

Samples of the parameter $X_{max}$ obtained from the CONEX simulations are considered. A Gaussian uncertainty of 
$\sigma[X_{max}]=20$ g cm$^{-2}$ and $\mu=0$ is assumed in order to take into account the detector response and the 
reconstruction method. Therefore, the distribution function of the reconstructed $X_{max}$ is given by Eq. (\ref{fdef}) 
with $g(X_{max})$ the distribution function corresponding to the physical fluctuations and $h(X)=G(X;0,\sigma[X_{max}])$,
a Gaussian distribution of mean value $\mu = 0$ and $\sigma = \sigma[X_{max}]$, which takes into account the response 
of the detectors and reconstruction methods. 

Four sets of 100 samples are considered. Each set of samples is noted as $S_{(M,m)}$ where $M$ indicates the independent 
values of $X_{max}$ (obtained from CONEX) in each sample and $m$ the number of repetitions of each shower, i.e., 
the number of times that the Gaussian distribution $h(X)=G(X;X_{max}^i,\sigma[X_{max}])$ is sampled for each of the $M$ 
independent values $X_{max}^i$ in each individual sample. Therefore, $S_{(110,1)}$, $S_{(10,11)}$, $S'_{(110,1)}$ 
and $S_{(22,5)}$ are considered, where $S_{(110,1)}$ and $S'_{(110,1)}$ just differ in the different values obtained
from the Gaussian distribution performed to include the detector response and reconstruction method. The number of 
events in each sample, belonging to the different sets, is $N_{ev}=M \times m=110$, the same for all kind of samples 
considered. 
 
\begin{figure}[!bt]
\begin{center}
\includegraphics[width=3.2in]{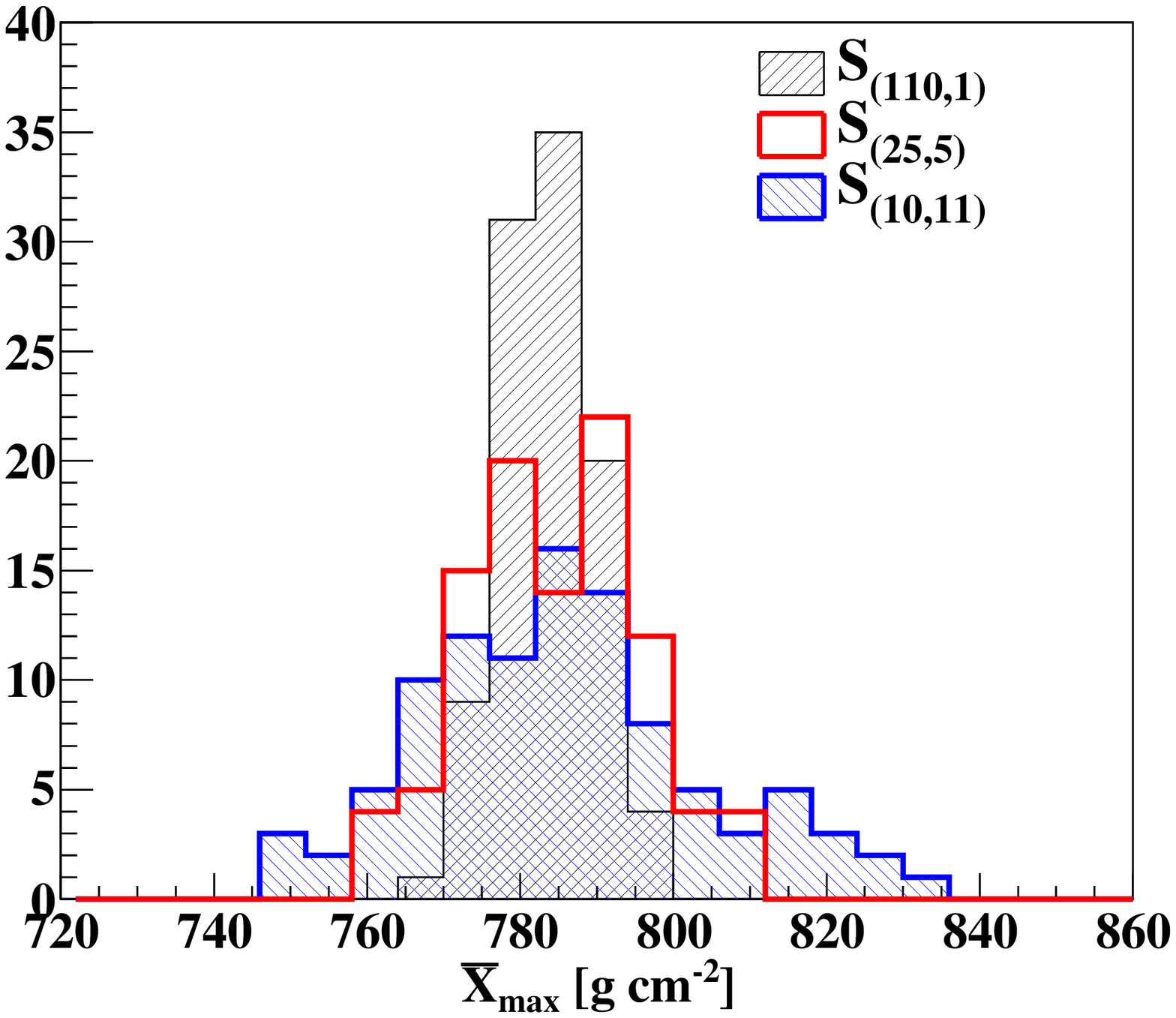}
\includegraphics[width=3.2in]{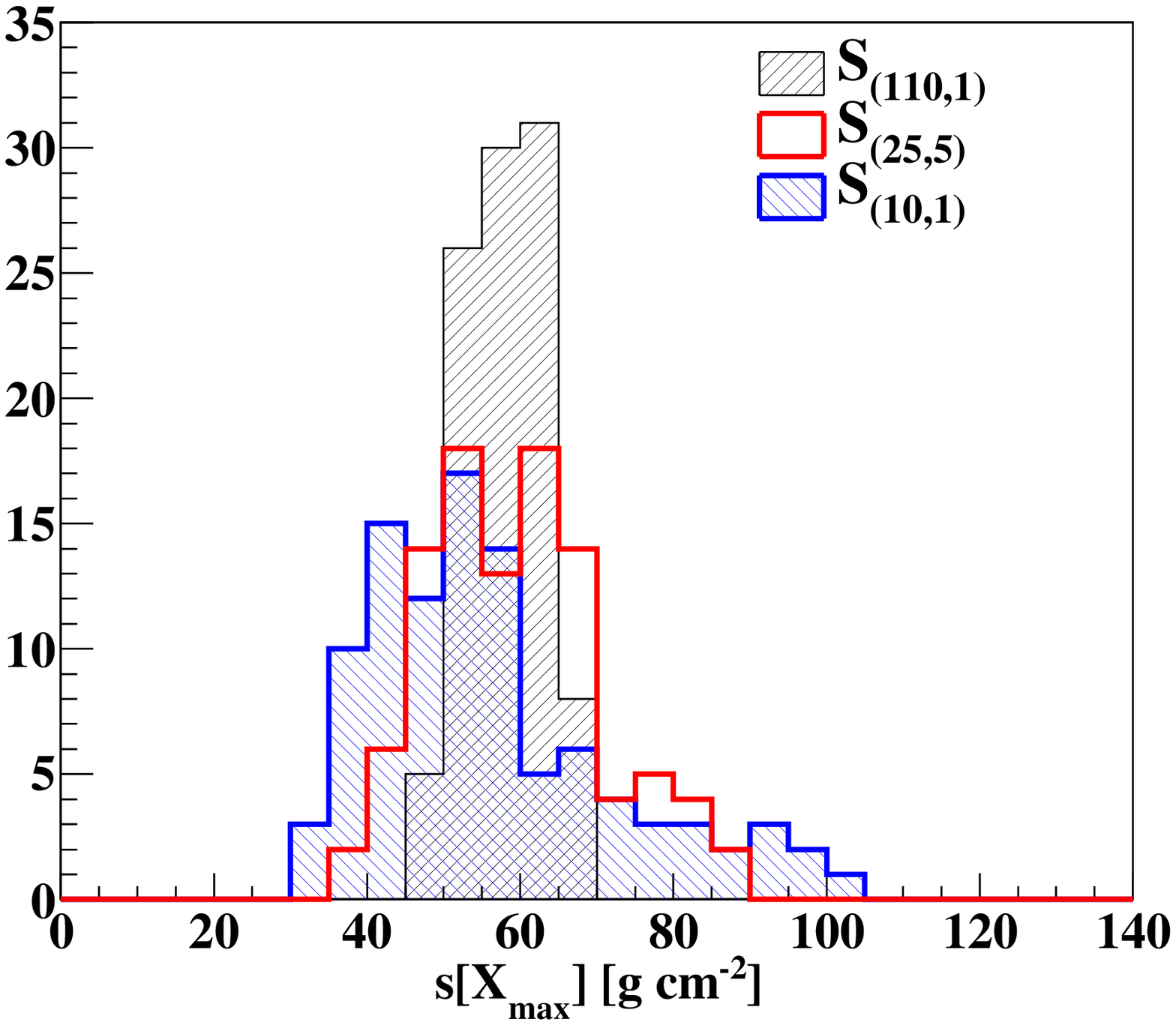}
\caption{Distributions of $\bar{X}_{max}$ and $s[X_{max}]$ for the different sets of samples considered.
\label{MeanRMS}}
\end{center}
\end{figure}

Figure \ref{MeanRMS} shows the distributions of the estimators of the average, $\bar{X}_{max}$, and the standard 
deviation, $s[X_{max}]$, for the sets of samples considered. It can be seen that, as expected, when the repetitions 
are included, the fluctuations increase and when the number of independent showers increases the fluctuations decrease.  
Figure \ref{MeanRMS} also shows that, although the distributions of $s[X_{max}]$ with repetitions have a tail towards 
larger values of grammage, which is not present in the corresponding without repetitions, the bias is not statistically 
significative. This is consistent with Eq. (\ref{Evarxrep}) which shows that the expected bias introduced by repetitions 
in the variance is proportional to $(m-1)/M m \cong 0.1$ for $S_{(10,11)}$.     

In order to illustrate the effects of repetitions on the density estimators, one-dimensional Gaussian kernels 
are used to estimate the density function of $X_{max}$. An adaptive bandwidth method, introduced by B. Silverman 
\cite{Silvermann:86}, is used to obtain better estimates of the density function.

\begin{figure}[!bt]
\begin{center}
\includegraphics[width=3.2in]{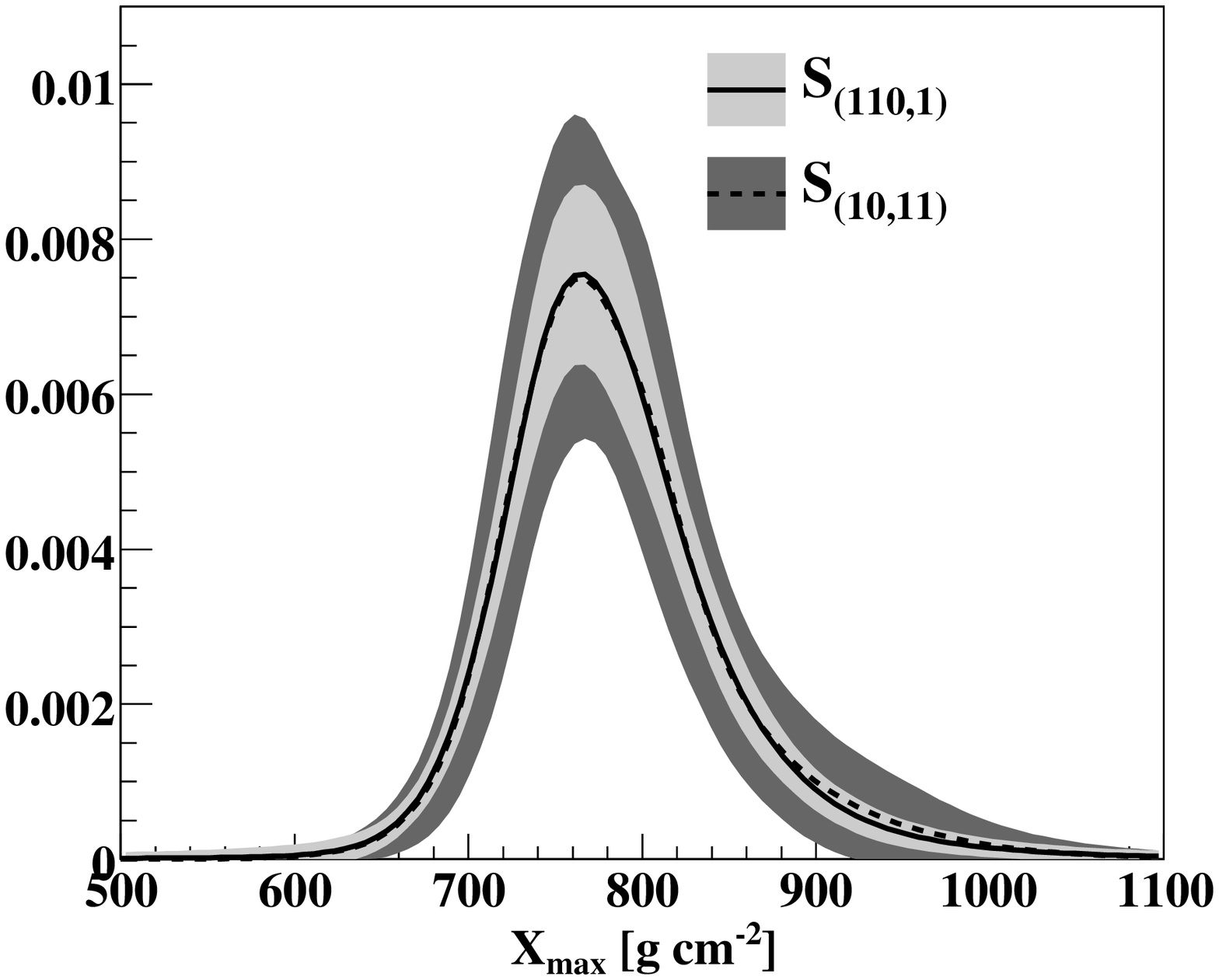}
\includegraphics[width=3.2in]{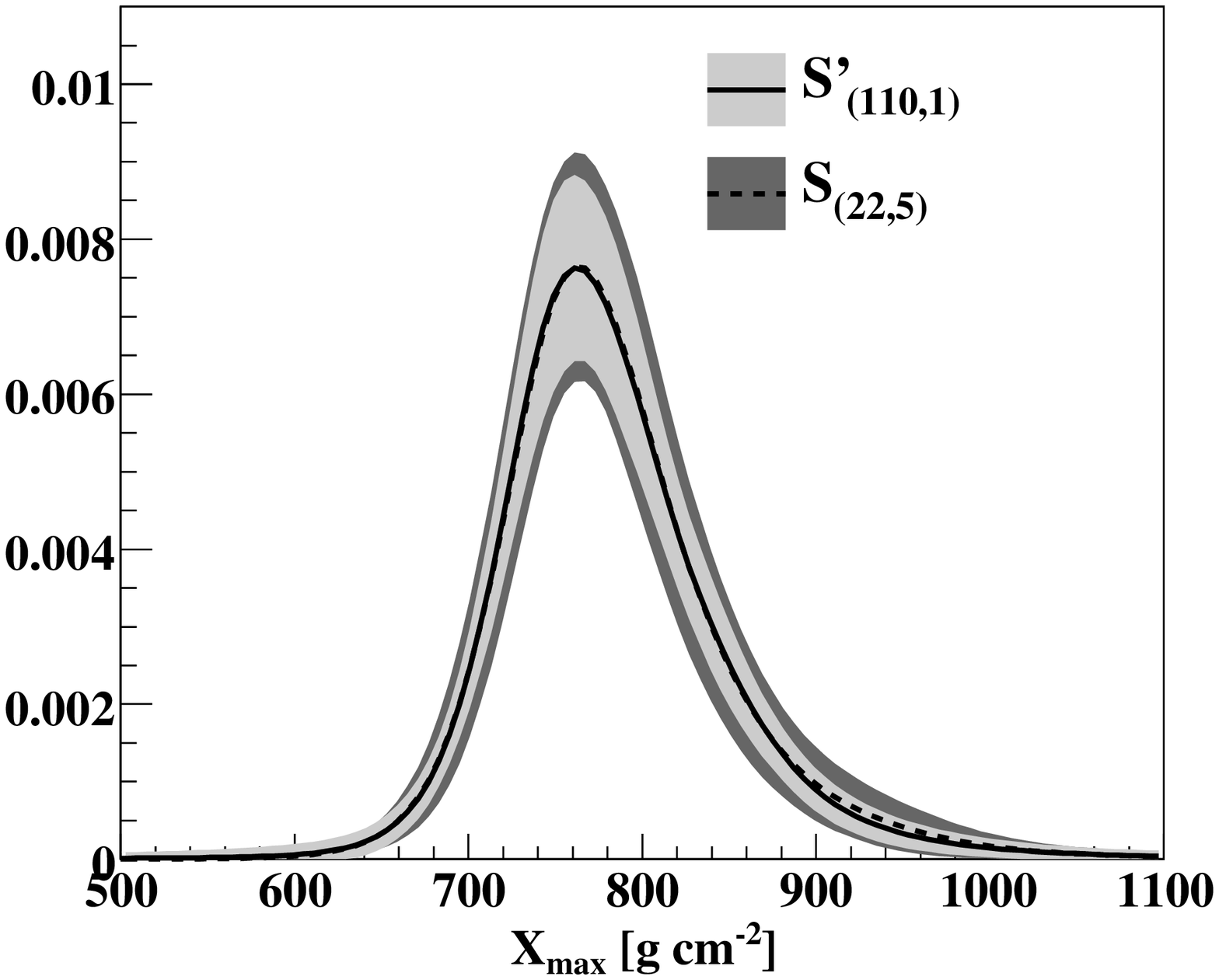}
\caption{Mean and one sigma regions for the density estimates obtained from the different samples considered. Darker
regions and dotted lines correspond to samples including multiple repetitions. 
\label{Estimates}}
\end{center}
\end{figure}

For each sample belonging to a given set a density estimate is obtained, therefore, 110 density estimates are
obtained for each set of samples considered. Figure \ref{Estimates} shows the mean value and the one sigma region 
obtained from the density estimates of each set. It can be seen that the mean values corresponding to samples with 
or without repetitions are very similar, which is consistent with the result obtained in subsection \ref{DensEst}. 
Also, as expected from Eq. (\ref{VarRep}), the fluctuations corresponding to sets including repetition are larger
and comparing the results obtained for $S_{(10,11)}$ and $S_{(22,5)}$ we see that the fluctuations in the latter 
case are smaller due to the smaller number of repetitions.  

\section{Conclusions}

In this work we present a study of the effects of recycling individual cosmic ray showers to simulate 
the detector response, which is a common practice in Monte Carlo simulations at the highest energies. We 
find that the standard estimators of the expected value, variance and covariance are modified. In particular, 
the average remains as a non-biased estimator of the expected value but the fluctuations are increased. For 
the standard estimators of the variance and covariance a bias proportional to $(m-1)/Mm$ appears when 
repetitions are included. Besides, as in the case of the average, the fluctuations of both estimators are 
increased. Finally, we study the effects introduced by repetition in the kernel density estimators obtained 
from finite samples. We find again that the expected value of the estimator is unchanged, i.e., the bias takes 
the same form. However, the pointwise fluctuations are increased and become more important as the ratio 
$(m-1)/Mm$ increases. 

\section*{Acknowledgements}
This work is partially supported by the Mexican agencies CONACyT, UNAM's CIC, and PAPIIT.
ADS is supported by a postdoctoral grant from the UNAM.

\end{document}